\documentstyle[11pt,newpasp,twoside]{article}
\def\edcomment#1{\iffalse\marginpar{\raggedright\sl#1\/}\else\relax\fi}
\reversemarginpar

\begin{document}
\title{Recent developments of inverse Compton scattering
model of pulsar radio emission}

\author{G. J. Qiao$^{1,2}$,
        R. X. Xu$^{1,2}$,
        J. F. Liu$^{1,2}$,
        B. Zhang$^{1,2}$,
        J. L. Han$^{1,3}$}
\affil{$^1$Beijing Astrophysics Center, CAS-PKU, Beijing 100871, China\\ 
$^2$Astronomy Department, Peking University, Beijing 100871, China\\
$^3$National Astronomical Observatories, CAS, Beijing 100012, China}

\begin{abstract}
Many theoretical efforts were made to understand the core and
conal emission identified from observations by Rankin (1983)
and Lyne and Manchester (1988). One of them, named as inverse
Compton scattering (ICS) model (Qiao \& Lin 1998), has
been proposed. It is found in the model that: there are central or
`core' emission beam, and one or two hollow conical emission
beams; the different emission components are emitted at different
heights; owing to different radiation components emitted from
different height, the observed emission beams can be shifted from
each other due to retardation and aberration effects; the sizes of
emission components change with frequencies. Recent developments
of the model include: simulations of pulse profiles at different
frequencies; studying the basic polarization properties of inverse
Compton scattering in strong magnetic fields; computing the
polarizations and spectrum of core and cones. A new classification
system was also proposed. The main results calculated from the
model are consistent with the observations.
\end{abstract}

\section{Introduction}

The emission beams of a radio pulsar have been identified as two (core,
inner conal, Lyne \& Manchester 1988) or three parts
(plus an outer conal, Rakin 1983) through
careful studies of the observed profiles and polarization
characteristics. Many theoretical models can only explain the
hollow cone beam. It is necessary to understand
the core emission theoretically. Some theoretical efforts have been
made, one of them is inverse Compton scattering model (ICS) model, 
which can get
both core and conal emission beams (Qiao and Lin 1998,
Liu et al. 1999; Qiao et al. 1999b; Xu et al. 1999a).

Up to now, following issues have been investigated for the model:

\noindent
(1). Inner gap structure and the explanation of some
phenomena (Zhang \& Qiao

1996; Qiao \& Zhang 1996; Zhang et al. 1997a,b), such as
mode-changing,

nulling.

\noindent
(2). Emission beams and emission regions of radio pulsars
(Qiao \& Line 1998).

\noindent
(3). Frequency behaviour of pulse profiles (Liu \& Qiao 1999;
 Qiao et al. 1999b).

\noindent
(4). The polarization properties of the ICS model in strong magnetic
fields (Xu

et al. 1999a);

\noindent
(5). Depolarization and position angle jumps (Xu et al. 1997);

\noindent
(6). Coherent ICS process in magnetosphere (Liu et al.
1999; Xu et al. 1999a).

\section{Basic idea of the ICS model}

The basic idea of the model can be found in
Qiao \& Lin (1998),  Qiao et al. (1999b), Xu et al. (1999a)
and Liu et al. (1999).  In the model, low frequency electromagnetic
waves are supposed to be produced
near the star surface due to the violent breakdown of
RS type vacuum gap (Ruderman \& Sutherland 1975). The waves 
are assumed to propagate freely in pulsar
magnetospheres and inverse Compton scattered by the
secondary particles produced in gap sparking processes.
The upscattered photons are in the radio band,
i.e., the observed radio emission.
With the simple dipole field,
the incident angle of the ICS decreases first, and then starts to
increase above a critical height.
The Lorentz factor of the secondary particles,
however, keeps decreasing due to various energy loss mechanisms
(mainly the ICS with the thermal photons near the surface). The
combination of the above two effects naturally results in the feature
that on a given field line, the emission has the same frequency
at three heights, corresponding to one core and two conal emission
components.

One basic ingredient of the ICS model is the vacuum gap, which has
been opposed by binding energy calculations. However, the idea that
pulsars are bare strange stars can solve the binding energy problem
completely (Xu \& Qiao 1998; Xu et al. 1999b).
If the vacuum gap could be formed,
the ICS of the primary particles off
the thermal photons actually take an important role in the inner gap
physics, both within and above the gap (Zhang \& Qiao 1996;
Zhang et al. 1997a,b). The energy loss behaviour of the secondary
particles is also influenced by ICS process (Zhang et al. 1997b).

\section{Emission beams and their properties}

The central or `core' emission beam, inner cone and outer cone
beams have been simulated
in the ICS model. We found that:

(1). `Core' emission should be a small hollow cone in fact, which
can be identified from de-composited Gaussian components
(Qiao et al. 1999a).

(2). Different emission components are emitted at different heights:
`core' emission is emitted at a place near the surface, `inner cone'
at a higher place, and `outer cone' at the highest.  Due to the
retardation and aberration effects caused by different heights,
polarization position angle can have two or three modes
at a given longitude (Xu et al. 1997).

(3). The beam size changes with frequencies.
As observing frequency increases, the `core' emission beam becomes
narrow, the `inner cone' becomes slightly wider or has little change,
and the `outer cone' also becomes narrow (Qiao \& Lin 1998).

\section{Classification and frequency behaviour of pulse profiles}

Based on the ICS model and the multi-frequency observations,
radio pulsars could be devided into two categories:

{\it Type I:} Pulsars with core and inner cone.
These pulsars has a shorter period $P$, and their
polar caps are larger (e.g. Fig.6b in Qiao \& Lin 1998). As the
impact angle of the line of sight gradually increases,
they are grouped into two sub-types, namely Ia (e.g. PSR
B1933+16) and Ib (e.g. PSR B1845-01).

{\it Type II:} Pulsars with all three parts of beam.
These pulsars have normal periods, and
the low frequency waves should be strong enough at high altitudes
to produce the radio emission. They can be further
grouped into three sub-types as the impact angle gradually
increases. Type IIa: pulsars
with five or six components at most observing frequencies
when the line of sight cut the core beam. The prototype
is PSR B1237+25.  Type IIb:
the impact angle is larger so that at higher frequencies the line-of-sight
missed the core beam. An example is PSR B2045-16. Type IIc:
The line of sight has the largest impact angle, so that only the
outer conical branch can be
observed. A typical pulsar is PSR B0525+21.

The profiles of most these types or subtypes have been simulated,
and typical examples were selected and compared its multi-frequency
observations (Qiao et al.1999b, Liu \& Qiao 1999).
Here are two examples.

{\it Type Ia:}
The multi-frequency observations of PSR B1933+16
show that it belongs to Type Ia pulsars.
It has a single component at low frequency, but becomes
triple at higher frequencies. Such behaviour can be simulated in
the model, since the radius of the
`inner' cone increases towards higher frequencies. 
This may be
an important feature of the ICS model distinguished from the other models.

{\it Type IIa:}
For such pulsars, we simulate a typical example, PSR B1237+25.
It is worth mentioning that the ICS model can interpret an
important characteristics of this pulsar: five components
in most frequency bands, but three components at very low frequencies.
This can hardly be explained by any other model.

\section{Polarization}

The polarization features of scattered emission by relativistic
electrons in the {\it strong} magnetic field were calculated from
the Stokes parameters of scattering emission (Xu et al. 1999a).
(1). When $\omega_{\rm in}\ll{eB\over mc}$, both $\omega_{\rm in}$
and $\omega_{\rm out}$ (the angular frequency of incident and
outgoing photons, respectively) are in radio band, the scattered
photons are {\it completely} linearly polarized, and its
polarization position angle is in the co-plane of the out-going
photon direction and the magnetic field.
(2). For resonant scattering at high energy bands, significant
circular polarization appears in the scattered emission. The
position angle of linear polarization is perpendicular to the
co-plane of out-going photon and the magnetic field, different
from the case in radio band.

The inverse Compton scattering of a bunch of particles outflowing
in pulsar magnetosphere should be coherent in order to
produce significant circular polarization for beamed radio
emission (Xu et al 1999a).
At a certain time an observer can only see a small part of an emission
beam radiated by a particle bunch, which we called `transient beam'.
(1). In ICS model, at a given frequency the transient beam have three parts 
(core, inner and outer cones), each of them is called `mini-beam', and
their polarization feature are quite different.
(2). Circular polarization is very strong (even up to 100\%) in the
core mini-beam and it is much less in the inner cone mini-beam.
(3). If the line of sight sweeps across the center of a core (or
inner conal) mini-beam, the circular polarization will experience a
central sense reversal, or else it will be dominated by one sense,
either the left hand or the right hand according to its traversing line
relative to the mini-beam.
(4). The position angles at a given longitude of {\it transient}
`sub-pulses' have diverse values around the projection of the
magnetic field. The variation range of position angles is larger
for core emission, but smaller for conal beam. When many such
`sub-pulses' from one mini-beam is summed up, the mean position
angle at the given longitude will be averaged to be the central
value, which is determined by the projection of magnetic field
lines.
(5). Stronger circular polarization should be observed in
sub-pulses with higher time resolution according to our model.

\section{Conclusion}

Besides the natural appearance of core and conal components, some
observational properties of radio pulsars can also be explained in
the ICS model, such as the frequency-dependent pulse profiles, the
polarization nature of mean pulses and individual pulses. We propose
here a classification method for grouping pulsar integrated pulses,
which may help to understand the multi-frequency pulse data.

\begin{acknowledgments}
This work is supported by National Nature Sciences Foundation of
China, by the Climbing project of China, and by the Youth
Foundation of PKU.
\end{acknowledgments}

\end{document}